# A Framework for Predictive Analysis of Stock Market Indices – A Study of the Indian Auto Sector

Jaydip Sen
Calcutta Business School, Diamond Harbor Road, Bishnupur – 743503
West Bengal, INDIA
email: jaydip.sen@acm.org

and

Tamal Datta Chaudhuri
Calcutta Business School, Diamond Harbour Road, Bishnupur – 743503
West Bengal, INDIA
email: tamalc@calcuttabusinessschool.org

**ABSTRACT**

Analysis and prediction of stock market time series data has attracted considerable interest from the research community over the last decade. Rapid development and evolution of sophisticated algorithms for statistical analysis of time series data, and availability of high-performance hardware has made it possible to process and analyze high volume stock market time series data effectively, in real-time. Among many other important characteristics and behavior of such data, forecasting is an area which has witnessed considerable focus. In this work, we have used time series of the index values of the Auto sector in India during January 2010 to December 2015 for a deeper understanding of the behavior of its three constituent components, e.g., the trend, the seasonal component, and the random component. Based on this structural analysis, we have also designed five approaches for forecasting and also computed their accuracy in prediction using suitably chosen training and test data sets. Extensive results are presented to demonstrate the effectiveness of our proposed decomposition approaches of time series and the efficiency of our forecasting techniques, even in presence of a random component and a sharply changing trend component in the time-series.

## 1. INTRODUCTION

Management theory has always emphasized that for any strategic decision, scanning the external environment is extremely important. A company does not exist in isolation. It is part of an industry, which itself is embedded in the domestic economy. In today's globalized environment, the latter, again, is not immune to world economic movements. Concepts like SWOT analysis, BCG Matrix, Kaplan's Balanced Score Card and Porter's Five Forces have made understanding the external environment of a company an essential part of evaluation. Without analyzing the sector within which the company operates, Strategic Management is not possible. It would thus be fair to say that for portfolio management, buying stocks of a company requires an analysis of the sector in which the company belongs.

Our focus on the performance of a sector is restricted to understanding whether it has strong seasonal characteristics, or whether it has a dominant trend factor, or whether its performance is random in nature. It is perfectly possible for a sector to display, dominantly, any of the above three characteristics, at different points of time. It is important to monitor these patterns continuously for both portfolio choice. Besides providing a framework for decomposition of time series data, for monitoring the performance of a portfolio, we have also designed approaches for forecasting and also computed their accuracy in prediction using suitably chosen training and test data sets.

## 2. OBJECTIVE OF THE STUDY

The purpose of this paper is to breakdown time series data of sectoral indices into trend, seasonal and random components. This will help in understanding the sector in the following ways. First, it will indicate the overall trend of the sector and hence help us understand whether the stocks of this sectors are short term or long term buys. Second, if seasonality patterns can be seen, then during which month which sector and hence which stock should be a good buy, can be inferred. Third, sectors, and hence stocks with dominant random patterns, can be used for pure speculative gains.

The second objective of the paper is to provide a framework for prediction purposes. It is important not only to understand a sector, but also to predict its performance in future. This is where we provide further insight into the Efficient Market Hypothesis. In this paper we perform this exercise for the Auto Sector in India.

The rest of the paper is organized as follows. Section 3 briefly discusses the methodology in constructing various time series and decomposing the time series into its components. Section 4 presents the results of decomposition of the auto sector index time series values into its trend, seasonal and random components. Inferences are made on the roles played by the three components in the overall time series index values. Section 5 presents five forecasting approaches and one approach for analyzing the behavior of the structural components of the auto sector index time series. Section 6 presents some related work in the current literature. Finally, Section 7 concludes the paper.

## 3. METHODOLOGY

In this work, we use the daily index data for the Auto sector for the period January 2010 to December 2015. The daily index values are first aggregated into monthly averages resulting into 70 values in the time series data. We use the **ts( )** function in the **TTR** library in the R programming language to convert the raw data into a monthly time series. The time series in R now is an aggregation of three components: (i) Trend, (ii) Seasonal, and (iii) Random. In order to make further investigations into the behavior of the time series data, we decompose the time series into its three components. For this purpose, we use the **decompose( )** function defined in the **TTR** library in R. After the decomposition, the components of the time series are studied in greater details so as to understand the behavior of the time series more closely. We also apply some robust forecasting techniques on this data and critically analyze the accuracy of each of the forecasting methods that we have applied.

This work is an extension of our earlier work [1]. In this extended work, we have proposed two more robust techniques of forecasting and have also compared the performances of different approaches of forecasting. We have also critically analyzed the reasons why some approaches have performed better in comparison with other approaches.

## 4. DECOMPOSITION RESULTS

In this Section, we present the results that we have obtained in time series decomposition work. We particularly focus on the Auto sector time series values and discuss the results that we have obtained from its decomposition.

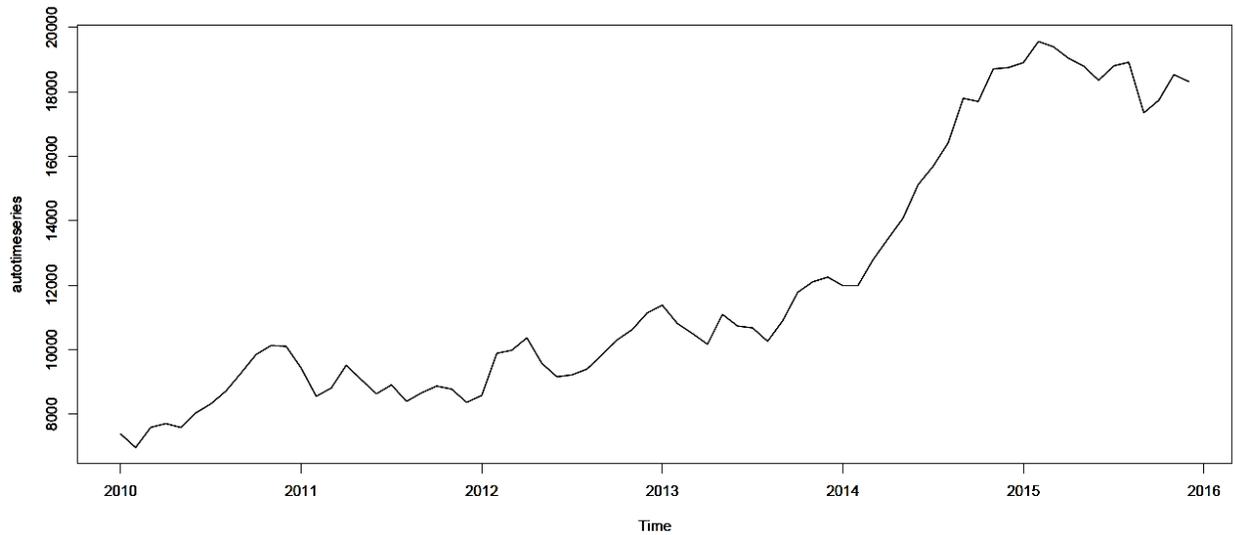

**Figure 1: Auto index time series (Jan 2010 – Dec 2015)**

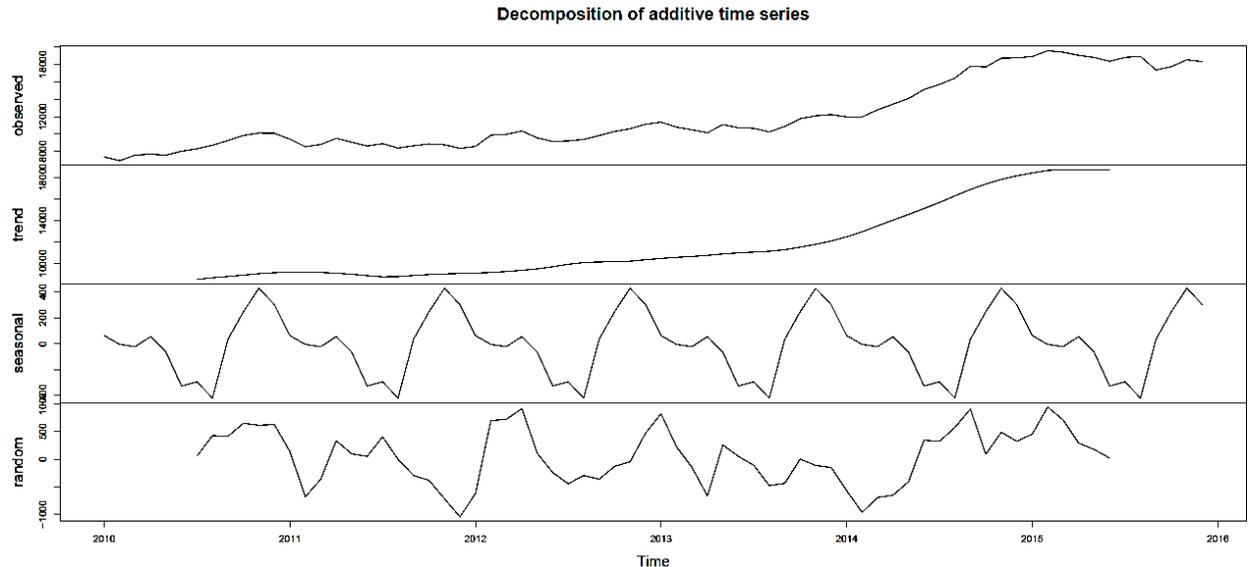

**Figure 2: Decomposition of auto index time series into trend, seasonal and random components**

**Figure1** shows the overall time series for the Auto sector index for the period January 2010 – December 2015. It is not difficult to see that the time series had an increasing trend till the curve exhibited a small downward fall during the latter part of 2015. **Figure 2** shows the decomposition results of the time series of **Figure 1**. The three components of the time series are shown separately so that their relative behavior can be visualized.

**Table 1** presents the numerical values of the time series data and its three components. The trend and the random components are not available for the period January 2010 – June 2010 and also for the period July 2015 – December 2015. This is due to the fact that trend computation requires long term data. In order to compute trend figures for January 2010 – June 2010 we need time series data from July 2009 – December 2009 (which is not available in our dataset). In the same line, for computing trend figures for July 2015- December 2015, time series data from January 2016 – June 2016 are needed. Due to the non-availability of the trend values for these periods, it is not possible to compute the random components too. Since the aggregate of the trend, seasonal and random components is the final time series figure and because of the fact that seasonal components remain constant for the same month over the period, the absence of trend values makes it impossible for us to compute the random components for these specific months.

**Table 1: Auto Sector Index Time Series and its Components (Jan 2010 – Dec 2015)**

| Year | Month | Time Series Aggregate | Trend | Seasonal | Random |
|---|---|---|---|---|---|
| **2010** | January | 7380 | | 64 | |
| | February | 6958 | | -5 | |
| | March | 7584 | | -23 | |
| | April | 7702 | | 55 | |
| | May | 7581 | | -65 | |
| | June | 8034 | | -326 | |
| | July | 8315 | 8552 | -294 | 56 |
| | August | 8710 | 8704 | -420 | 426 |
| | September | 9269 | 8821 | 35 | 413 |
| | October | 9844 | 8947 | 247 | 649 |
| | November | 10127 | 9085 | 429 | 614 |
| | December | 10100 | 9171 | 302 | 627 |
| **2011** | January | 9426 | 9220 | 64 | 142 |
| | February | 8547 | 9231 | -5 | -679 |
| | March | 8806 | 9192 | -23 | -364 |
| | April | 9515 | 9126 | 55 | 334 |
| | May | 9061 | 9029 | -65 | 97 |
| | June | 8626 | 8900 | -326 | 52 |
| | July | 8902 | 8792 | -294 | 404 |
| | August | 8390 | 8812 | -420 | -2 |
| | September | 8656 | 8916 | 35 | -295 |
| | October | 8866 | 9001 | 247 | -382 |
| | November | 8771 | 9057 | 429 | -715 |
| | December | 8359 | 9100 | 302 | -1044 |

|      | Month | | | | |
|------|-----------|-------|-------|------|------|
| **2012** | January | 8576 | 9135 | 64 | -623 |
|      | February | 9883 | 9190 | -5 | 698 |
|      | March | 9979 | 9281 | -23 | 720 |
|      | April | 10363 | 9390 | 55 | 917 |
|      | May | 9568 | 9527 | -65 | 106 |
|      | June | 9154 | 9720 | -326 | -240 |
|      | July | 9215 | 9953 | -294 | -444 |
|      | August | 9394 | 10108 | -420 | -294 |
|      | September | 9841 | 10168 | 35 | -362 |
|      | October | 10299 | 10182 | 247 | -130 |
|      | November | 10620 | 10237 | 429 | -46 |
|      | December | 11139 | 10366 | 302 | 471 |
| **2013** | January | 11379 | 10492 | 64 | 823 |
|      | February | 10809 | 10589 | -5 | 225 |
|      | March | 10499 | 10669 | -23 | -147 |
|      | April | 10164 | 10774 | 55 | -666 |
|      | May | 11091 | 10897 | -65 | 259 |
|      | June | 10731 | 11005 | -326 | 51 |
|      | July | 10672 | 11077 | -294 | -111 |
|      | August | 10255 | 11151 | -420 | -476 |
|      | September | 10893 | 11295 | 35 | -437 |
|      | October | 11776 | 11527 | 247 | 2 |
|      | November | 12103 | 11787 | 429 | -113 |
|      | December | 12247 | 12095 | 302 | -150 |
| **2014** | January | 11983 | 12487 | 64 | -567 |
|      | February | 11985 | 12952 | -5 | -962 |
|      | March | 12783 | 13497 | -23 | -691 |
|      | April | 13437 | 14031 | 55 | -650 |
|      | May | 14078 | 14554 | -65 | -411 |
|      | June | 15118 | 15100 | -326 | 344 |
|      | July | 15688 | 15660 | -294 | 322 |
|      | August | 16418 | 16264 | -420 | 574 |
|      | September | 17798 | 16855 | 35 | 908 |
|      | October | 17700 | 17364 | 247 | 88 |
|      | November | 18712 | 17795 | 429 | 488 |
|      | December | 18752 | 18126 | 302 | 324 |
| **2015** | January | 18907 | 18391 | 64 | 452 |
| | February | 19565 | 18625 | -5 | 945 |
| | March | 19397 | 18711 | -23 | 709 |
| | April | 19041 | 18693 | 55 | 292 |
| | May | 18799 | 18688 | -65 | 176 |
| | June | 18357 | 18662 | -326 | 21 |
| | July | 18806 | | -294 | |
| | August | 18918 | | -420 | |
| | September | 17348 | | 35 | |
| | October | 17738 | | 247 | |
| | November | 18535 | | 429 | |
| | December | 18317 | | 302 | |

**Observations:**

1. From **Table 1**, we observe that the seasonal components for the Auto sector indices are positive during the period September-January, with the highest value occurring in the month of November. The seasonal component is the minimum in the month of June every year. The trend values consistently increased over the period 2010 – 2015. However, the rate of growth of the trend value has decreased during January 2015- June 2015 and the decreasing trend might have continued possibly even after that period. The random component has shown considerable fluctuations in its values. However, the trend being the predominant component in the overall time series, the time series is quite amenable for forecasting.

2. It is natural for the auto sector to have a dominant seasonal component as purchase of vehicles coincide, both with the religious festivals, and also during the third quarter, as during this period the economic activity starts to pick up. The results from the agricultural sector add to the seasonality as the impacts of the rains are available from the month of September onwards.

## 5. RESULTS OF FORECASTING

In this Section, we present some forecasting methods that we have applied on the time series data of the Auto sector index. We have proposed five different approaches in forecasting and also have presented their relative performance in term of their forecast accuracies. However, the main observation from our analysis is that the Auto sector index is very much dominated by its trend and the seasonal components with a weak random component playing a minor role.

The five different methods for forecasting, and a method for understanding the strengths of the trend and the seasonality components in the time series data are discussed as bellow in this Section.

**Method I:** The time series data of the Auto sector index from January 2010 to December 2014 are used for forecasting the monthly indices for the year 2015. The forecasting is made at the end of December 2014. Error in forecasting is also computed for each month in order to have an idea about the accuracy in forecasting technique.

**Method II:** Forecasting for the monthly indices for the year 2015 is made on the basis of time series data from January 2010 till the end of the previous month for which the forecast is made. For example, for the purpose of forecasting the monthly index for March 2015, time series data from January 2010 till February 2015 are considered. As in Method 1, error in forecasting is also computed.

**Method III:** In this method, we first use the time series data for the Auto sector monthly indices from January 2010 to December 2014 to compute its trend and seasonal components. This method yields the trend component from July 2010 to June 2014 with the trend values for the first six months and last six months being truncated. Based on the trend values till June 2014, we make forecasts for the trend values for the period January 2015 to June 2015, using the **HoltWineters( )** function in R with a changing trend and a seasonal component. The forecasted trend values are added to the seasonal components values of the corresponding months (based on the time series data from January 2010 to December 2014) to arrive at the forecasted aggregate of the trend and seasonal components. Now we consider the full time series of the Auto sector indices from January 2010 to December 2015 and decompose it into its trend, seasonal and random components. We compute the aggregate of the actual trend and the actual seasonal component values for the period January 2015 to June 2015. Finally, to have an idea about forecasting accuracy, we compute the percentage of deviation of the actual aggregate of trend and seasonal component values with their corresponding forecasted aggregate values for each month during January 2015 to June 2015.

**Method IV:** We use **Auto Regressive Integrated Moving Average (ARIMA)** based approach of forecasting in this method. For the purpose of building the ARIMA model, we use the Auto sector time series data for the period January 2010 – December 2014. Based on this training data set, we compute the three parameters of the **Auto Regressive Moving Average (ARMA)** mode, i.e. the **Auto Regression** parameter (p), the **Difference** parameter (d), and the **Moving Average** parameter (q). The values of the three parameters are used to develop the ARIMA model for the purpose of forecasting. Finally, the ARIMA model is used to predict the time series values for all the months in the year of 2015. Since forecasts for all the months of 2015 are made at the end of December 2014, the prediction horizon for the ARIMA model in this approach is 1 year.

**Method V:** In this approach, we use ARIMA model with a forecast horizon of one month. Hence, for the purpose of prediction, the training data set for the ARIMA model contained time series data form January 2010 till the last month for which the forecast was made. For example, for the purpose of prediction for the month of May 2015, the time series data form January 2010 till April 2015 were considered. Since the training data set for the ARIMA model constantly changed, we evaluated the ARIMA parameters every time before the forecasting is made for each month of 2015.

**Method VI:** The objective of this approach is to have an insight about how strong is the presence of the trend and seasonality in the auto sector index time series. In this method, we first consider the time series of the Auto sector month indices during January 2010 to December 2014. The time series is decomposed into its trend, seasonal and random components and we compute the aggregate of the trend and the seasonal components during July 2010 to June 2014. Note that the trend components from January 2010 to June 2010 and also from July 2014 to December 2014 are not available after the decomposition. Next, we consider the time series data from January 2011 to December 2015. We again compute the aggregate of the trend and the seasonal components for the new time series (i.e., the time series from January 2011 to December 2015). In order to have an idea about the change in the aggregate values of the trend and the seasonal components, we compute the percentage of deviation of the computed aggregate of the trend and the seasonal components for each month during June 2011 to July 2014, computed based on the two time series (January 2010 – December 2014 and January 2011 – December 2015).

**RESULTS**

**Method I:** As mentioned earlier in this Section, we make forecast for each month of 2015 based on time series data from January 2010 to December 2014. We use *HoltWinters* function in R library "forecast" for this purpose. In order to make a robust forecasting, we use HoltWinters model with a varying trend and an additive seasonal component that best fits the Auto index time series data. The forecast "horizon" in the *HoltWinters* model has been chosen to be 12 so that the forecasted values for all months of 2015 can be obtained. The results obtained of this method are presented in **Table 2**.

**Table 2: Computation Results using Method I**

| Month | Actual Value | Forecasted Value | Error Percentage |
|---|---|---|---|
| (A) | (B) | (C) | (C-B)/B *100 |
| Jan | 18907 | 18507 | 2.11 |
| Feb | 19565 | 17988 | 8.05 |
| Mar | 19397 | 18384 | 5.22 |
| Apr | 19041 | 19365 | 1.70 |
| May | 18799 | 19546 | 3.97 |
| Jun | 18357 | 19544 | 6.47 |
| Jul | 18806 | 19873 | 5.67 |
| Aug | 18918 | 20350 | 7.57 |
| Sep | 17348 | 21222 | 22.33 |
| Oct | 17738 | 21927 | 23.62 |
| Nov | 18535 | 22487 | 21.32 |
| Dec | 18317 | 22686 | 23.85 |

**Observations:** We observe from Table 2 that the forecasted values closely match the actual values even when the forecast horizon is long (12 months). This clearly shows that *HoltWinters* model with changing trend and additive seasonal components is very effective in forecasting Auto sector monthly indices during the period 2010 -2015. The error values during September – December 2015 are comparatively larger due to sudden downward movement in the index. The sudden decrease in the time series values from September 2015 has caused the trend values to decrease. This was impossible to predict in December 2014. Hence, the errors in forecasting are larger for the last four months of 2015.

**Method II:** As discussed earlier, in this approach we forecast the Auto sector index for each month in 2015 by taking into account time series data till the month before the month of forecast. We use *HoltWinters* model with additive seasonal component having a forecast horizon of 1 month. Since the forecast horizon is smaller, the model can capture any possible change in trend and seasonal components more effectively than Method I. The only factors that can induce error in forecasting are: (i) any appreciable change in the seasonal component, (ii) a very strong and abruptly changing random component. The results of this method are presented in **Table 3.**

**Table 3: Computation Results using Method II**

| Month | Actual Value | Forecasted Value | Error Percentage |
|---|---|---|---|
| (A) | (B) | (C) | (C-B)/B *100 |
| Jan | 18907 | 18507 | 2.12 |
| Feb | 19565 | 18426 | 5.82 |
| Mar | 19397 | 19825 | 2.21 |
| Apr | 19041 | 20307 | 6.65 |
| May | 18799 | 19687 | 4.72 |
| Jun | 18357 | 18937 | 3.16 |
| Jul | 18806 | 18609 | 1.05 |
| Aug | 18918 | 19077 | 0.84 |
| Sep | 17348 | 19794 | 14.10 |
| Oct | 17738 | 18038 | 1.70 |
| Nov | 18535 | 18104 | 2.33 |
| Dec | 18317 | 18470 | 0.84 |

**Observations:** We observe from **Table 3** that the forecasted values very closely match with the actual values. This clearly demonstrates that *HolWinters* additive model with a prediction horizon of 1 month can very effectively and accurately forecast future time series values. The high error value in the month of September is due to the sudden downward trend of the time series in that month. This was impossible to predict in August 2014 when the forecasting was done, as the downward fall in time series was very abrupt.

**Method III:** In the earlier part of this Section, we have already discussed the approach followed in this method. We have used the time series data of the Auto sector indices from January 2010 to December 2015 to compute the actual values of the trend and the seasonal components.

However, since the actual values of trend component are not available for the period July 2015 – December 2015, we concentrate only on the period January 2015 to June 2015 for the purpose of forecasting. The actual trend and seasonal component values and their aggregated monthly values are noted in Columns B, C and D respectively in **Table 4**. Now, using the time series data during January 2010 to December 2014, the trend and the seasonal components are recomputed. Since the trend values during July 2014 to December 2014 will not be available after this computation, we make a forecast for the trend values for the period January 2015 to June 2015 using *HoltWinters* forecasting model with a changing trend and an additive seasonal component. The forecasted trend values and the past seasonal component values and their corresponding aggregate values are noted in columns E, F and G respectively in **Table 4**. The error values are also computed.

**Observation:** The results obtained using Method III are presented in **Table 4**. We observe that the percentage error in forecast have consistently increased from a value of 8.36 to 24.73. However, this is expected as the forecast error usually increases with the increase in time horizon for forecast. Considering the fact that, the trend is forecasted over a period of one year (forecasting for January 2015 – June 2015 being done at the end of June 2014), and there has been a considerable change in the behavior of the time series during this period, it can be concluded that *HoltWinters* forecasting model with a changing trend and an additive seasonality component has performed reasonably well over a long horizon of forecasting.

**Table 4: Computation Results using Method III**

| Month | Actual Trend | Actual Seasonal | Actual (Trend + Seasonal) | Forecasted Trend | Past Seasonal | Forecasted (Trend + Seasonal) | % Error |
|---|---|---|---|---|---|---|---|
| A | B | C | D | E | F | G | (G-D)/D *100 |
| Jan | 18391 | 64 | 18455 | 20029 | 61 | 20090 | 8.86 |
| Feb | 18625 | -5 | 18620 | 20801 | -131 | 20670 | 11.01 |
| Mar | 18710 | -23 | 18687 | 21439 | -90 | 21349 | 14.30 |
| Apr | 18693 | 55 | 18748 | 21995 | 93 | 22088 | 17.82 |
| May | 18688 | -65 | 18623 | 22544 | 1 | 22545 | 21.06 |
| Jun | 18662 | -326 | 18336 | 23091 | -221 | 22870 | 24.73 |

**Method IV:** In this method, we have applied **Auto Regressive Integrated Moving Average (ARIMA)** technique for making forecasting on the Auto sector time series data. We have first used **auto.arima( )** function defined in the **forecast** package in R for identifying the values of the parameters for ARIMA function for the Auto time series. For this purpose, we have used the auto sector time series values for the period January 2010 – December 201. Applying **auto.arima( )** function on this time series, we have obtained the parameter values for the time series as: Auto Regression parameter (p) = 0, Difference parameter (d) = 2, Moving Average parameter (q) = 1. Therefore, the auto sector time series for the period January 2010 – December

2014 is designed as an Auto Regressive Moving Average (ARMA) model - ARMA (0, 2, 1). From this ARMA (0, 2, 1) model we construct the corresponding ARIMA model using the **arima( ) function** in R with the two parameters as: (i) auto sector time series, (ii) the order of the ARMA i.e., (0, 2, 1). Using the resultant ARIMA model, we call the function **forecast.Arima( )** with parameters: (i) the ARIMA model and the time horizon of forecast. The advantage of the ARIMA-based approach is that we can make forecast all the months of 2015 based on the time series values till December 2014. In this method (i.e., Method IV), we make the forecast for all the months of 2015 based on the time series values from January 2010 to December 2014, resulting in a forecast horizon of 12 months. We compare the forecasted values with the actual time series values for each month of 2015 and compute the error. The results are presented in **Table 5**.

**Observation:** The results obtained using Method IV are presented in **Table 5**. It may be noted that the error in forecast for all the months have been very moderate considering the fact that the forecast horizon was long (1 year). Moreover, the error had been very low till the month of August 2015. The increase in the error rate from September 2015 onwards is largely due to the fact that there was a sudden slump in the Auto sector index in September 2015 and that downward trend continued. Since the forecasting was made in December 2014, it was impossible for the ARIMA model to predict that sudden abnormal behavior of the trend that might have happened due to several exogenous economic parameters.

**Method V:** In this approach, we have used ARIMA model with a forecast horizon of one month. The methodology used for building the ARIMA model is the same as it was in Method IV. However, since we use a training data set that is constantly increasing in size, we re-evaluate the parameters of the ARIMA model every time we use it in forecasting. In other words, for each month of 2015, before we make the forecast for the next month, we compute the values of the parameters of the ARIMA model. The results obtained on application of this method are presented in **Table 6**.

**Observations:** The forecast errors in this method are very small as can be seen in **Table 6**. Since the forecast horizon is just one month, and the ARIMA method captures the dynamic behavior of the time series very effectively, even a strong presence of the a seasonality component and a changing trend in the time series cannot adversely affect the robustness of the forecasting method in this approach. For example, in spite of a sudden slump in the time series value in September 2015, the percentage error in forecasting using this method has been found to be only around 9.

**Table 5: Computation Results using Method IV**

| Month | Actual Value | Forecasted Value | Error Percentage |
|---|---|---|---|
| (A) | (B) | (C) | (C-B)/B *100 |
| Jan | 18907 | 18569 | -1.79 |
| Feb | 19565 | 18717 | -4.33 |
| Mar | 19397 | 18903 | -2.55 |
| Apr | 19041 | 19102 | 0.32 |
| May | 18799 | 19305 | 2.69 |
| Jun | 18357 | 19507 | 6.26 |
| Jul | 18806 | 19708 | 4.80 |
| Aug | 18918 | 19904 | 5.21 |
| Sep | 17348 | 20096 | 15.84 |
| Oct | 17738 | 20282 | 14.34 |
| Nov | 18535 | 20463 | 10.40 |
| Dec | 18317 | 20638 | 12.67 |

**Table 6: Computation Results using Method V**

| Month | Actual Value | Forecasted Value | Error Percentage |
|---|---|---|---|
| (A) | (B) | (C) | (C-B)/B *100 |
| Jan | 18907 | 18569 | -1.79 |
| Feb | 19565 | 18955 | -3.12 |
| Mar | 19397 | 19373 | -0.12 |
| Apr | 19041 | 19347 | 1.61 |
| May | 18799 | 18933 | 0.71 |
| Jun | 18357 | 18725 | 2.00 |
| Jul | 18806 | 18220 | -3.12 |
| Aug | 18918 | 18939 | 0.11 |
| Sep | 17348 | 18951 | 9.24 |
| Oct | 17738 | 16899 | -4.73 |
| Nov | 18535 | 18450 | -0.46 |
| Dec | 18317 | 18683 | 2.00 |

In **Table 7**, we have summarized the performance of the five forecasting approaches that we have discussed so far. For the purpose of comparison, we have chosen four metrics: (i) minimum (Min) error rate, (ii) maximum (Max) error rate, (iii) mean error rate, and (iv) standard deviation (SD) of error rates. As observed in **Table 7**, Method V that used ARIMA with a forecast horizon of one month has performed best in all the four metrics of performance. Method II that used HoltWineters forecasting methods with a forecast horizon of one month has been the next method in terms of its performance based on all the four metrics. Method IV is the third best performer. Although the mean error rate of Method IV has been somewhat adversely affected by a large value of its max error rate of 15.84, considering the fact that this method of ARIMA used a forecast horizon of one year, we consider its performance quite acceptable. Both Method I and

Method III has high mean error rate. However, Method I that used *HoltWinters* forecasting technique with a prediction horizon of one year has performed better in comparison to Method III that used forecasted value of the trend and the aggregate of the forecasted trend values over a period of one year with the past seasonal values to predict the aggregate of trend and the seasonal components for the period January 2015 – June 2015

**Table 7: Comparison of the Performance of the Forecasting Methods**

| Metrics / Methods | Min Error | Max Error | Mean Error | SD of Errors |
|---|---|---|---|---|
| Method 1 | 1.7 | 23.85 | 10.99 | 8.93 |
| Method II | 0.84 | 14.1 | 3.80 | 3.77 |
| Method III | 8.86 | 24.73 | 16.30 | 6.06 |
| Method IV | 0.32 | 15.84 | 6.77 | 5.23 |
| Method V | 0.11 | 9.24 | 2.41 | 2.55 |

**Method VI:** The objective of this method is to gain an insight into the contribution of the trend and the seasonal components of the time series on the overall auto sector aggregate index. As we mentioned earlier in this Section, this approach is based on comparison of the aggregate of the trend and the seasonal components of a time series over two different period of time. First, we construct a time series using the data for the period January 2010 to December 2014, and then compute the trend and the seasonal components and their aggregate values. We refer to this computation as **Computation 1**. The trend, the seasonal and their aggregate values in **Computation 1** are noted in columns A, B and C respectively in **Table 8**. Next, we construct the second time series using the data for the period January 2011 to December 2015 and repeat the computation of the trend, the seasonal and their aggregate values. We refer to this computation as **Computation 2**. The trend, the seasonal and their aggregate values in **Computation 2** are noted in columns D, E and F respectively in **Table 8**. The percentages of variation of the aggregate values in both computations are noted for each month for the period July 2011 to June 2014. If there is a structural difference between the time series data in 2010 and 2015, then we expect that difference to be reflected in the aggregate of the trend and the seasonal values.

**Table 8: Computation Results using Method VI**
**(Structural Analysis of Trend and Seasonal Components of the Auto Index Time Series for the Period: July 2014 – June 2015)**

| Year | Month | Computation 1 (Based on 2010 – 2014) | | | Computation 2 (Based on 2011 – 2015) | | | % Variation |
|------|-------|-------|-------|-------|-------|-------|-------|-------|
| | | Trend | Seasonal | Sum | Trend | Seasonal | Sum | |
| | | A | B | C = A + B | D | E | F = D + E | (F - C)/C *100 |
| 2011 | Jul | 8792 | -264 | 8498 | 8792 | -258 | 8534 | 0.36 |
| | Aug | 8812 | -452 | 8360 | 8812 | -477 | 8335 | -0.30 |
| | Sep | 8916 | -82 | 8834 | 8916 | -19 | 8897 | 0.71 |
| | Oct | 9000 | 336 | 9336 | 9000 | 134 | 9134 | -2.16 |
| | Nov | 9057 | 417 | 9474 | 9057 | 325 | 9382 | -0.97 |
| | Dec | 9100 | 332 | 9432 | 9100 | 195 | 9295 | -1.45 |
| 2012 | Jan | 9135 | 61 | 9196 | 9135 | 77 | 9212 | 0.17 |
| | Feb | 9190 | -131 | 9059 | 9190 | 214 | 9404 | 3.80 |
| | Mar | 9281 | -90 | 9191 | 9281 | 118 | 9399 | 2.26 |
| | Apr | 9390 | 93 | 9483 | 9390 | 21 | 9411 | -0.76 |
| | May | 9527 | 1 | 9528 | 9527 | -40 | 9487 | -0.43 |
| | Jun | 9720 | -221 | 9499 | 9720 | -289 | 9431 | -0.72 |
| | Jul | 9952 | -264 | 9688 | 9952 | -258 | 9694 | -0.06 |
| | Aug | 10108 | -452 | 9656 | 10108 | -477 | 9631 | -0.26 |
| | Sep | 10168 | -82 | 10086 | 10168 | -19 | 10149 | 0.62 |
| | Oct | 10181 | 336 | 10517 | 10181 | 134 | 10315 | -1.92 |
| | Nov | 10236 | 417 | 10653 | 10236 | 325 | 10561 | -0.86 |
| | Dec | 10366 | 332 | 10698 | 10366 | 195 | 10561 | -1.28 |
| 2013 | Jan | 10492 | 61 | 10553 | 10492 | 77 | 10569 | 0.15 |
| | Feb | 10582 | -131 | 10451 | 10589 | 214 | 10803 | 3.37 |
| | Mar | 10669 | -90 | 10579 | 10669 | 118 | 10787 | 1.97 |
| | Apr | 10774 | 93 | 10867 | 10774 | 21 | 10795 | -0.66 |
| | May | 10897 | 1 | 10898 | 10897 | -40 | 10857 | -0.37 |
| | Jun | 11005 | -221 | 10784 | 11005 | -289 | 10716 | -0.63 |
| | Jul | 11077 | -264 | 10813 | 11077 | -258 | 10819 | 0.06 |
| | Aug | 11151 | -452 | 10699 | 11151 | -477 | 10674 | -0.23 |
| | Sep | 11295 | -82 | 11213 | 11295 | -19 | 11276 | 0.56 |
| | Oct | 11527 | 336 | 11863 | 11526 | 134 | 11660 | -1.71 |
| | Nov | 11787 | 417 | 12204 | 11787 | 325 | 12112 | -0.75 |
| | Dec | 12095 | 332 | 12427 | 12095 | 195 | 12290 | -1.10 |
| 2014 | Jan | 12487 | 61 | 12548 | 12487 | 77 | 12564 | 0.13 |
| | Feb | 12952 | -131 | 12821 | 12952 | 214 | 13166 | 2.69 |
| | Mar | 13497 | -90 | 13407 | 13497 | 118 | 13615 | 1.55 |
| | Apr | 14031 | 93 | 14124 | 14031 | 21 | 14052 | -0.51 |
| | May | 14553 | 1 | 14554 | 14553 | -40 | 14513 | -0.28 |
| | Jun | 15100 | -221 | 14879 | 15100 | -289 | 14811 | -0.46 |

**Observation:** From **Table 8**, it is quite evident that the aggregate of the trend and the seasonal components had remained consistently the same over the period July 2011 to June 2014. This indicates that there has been no structural change in the time series from the year 2010 to the year 2015. It may be noted that we could not make computations during January 2011 to June 2011 and also during July 2014 to December 2014 due to non-availability of trend values during those periods. Since the change of the time series due to substitution of the 2010 data by 2015 data has virtually no impact on the trend and the seasonal components, we conclude that the impact of the random component is not significant, and the auto sector time series is quite amenable for robust forecasting.

## 6. RELATED WORK

Forecasting of daily stock prices has attracted considerable attention from the research community. Neural network based approaches have been proposed to make various kind of forecasting. Mostafa used neural network technique to predict stock market movements in Kuwait [2]. Kimoto et al applied neural network-based approach using historical accounting data and various macroeconomic parameters to forecast variations in stock returns [3]. Leigh et al used liner regression and simple neural network models for prediction stock market indices in the New York Stock Exchange for the period 1981-1999 [4]. Hammad et al have demonstrated that artificial neural network (ANN) model can be trained so that it converges while maintaining high level of precision in forecasting of stock prices [5]. Dutta et al used ANN models for forecasting Bombay Stock Exchange's SENSEX weekly closing values for the period of January 2002-December 2003 [6]. Ying et al used Bayesian network (BN)-approach to forecast stock prices of 28 companies listed in DJIA during 198-1998. Tsai and Wang showed results that highlighted the fact that Bayesian Network-based approaches have better forecasting ability than traditional regression and Neural Network-based approaches [7]. Tseng et al applied traditional time series decomposition (TSD), HoltWinters (H/W) models, Box-Jenkins (B/J) methodology and Neural Network- based approach to 50 randomly selected stocks from September 1, 1998 to December 31, 2010 with an objective of forecasting future stock prices [8]. They have observed that forecasting errors are lower for B/J, H/W and normalized Neural network model, while the errors are appreciably large for time series decomposition and non-normalized Neural Network model. Moshiri and Cameron [9] designed a Back Propagation Network (BPN) with econometric models to forecast inflation using (i) Box-Jenkins Autoregressive Integrated Moving Average (ARIMA) model, (ii) Vector Autoregressive (VAR) model and (ii) Bayesian Vector Autoregressive (BVAR) model. Datta Chaudhuri and Ghosh presented Artificial Neural Network (ANN) models based on various back propagation algorithms for the purpose of predicting volatility in the Indian stock market through volatility of NIFTY returns and volatility of Gold returns [10].

ANN and Hybrid systems are particularly effective in forecasting stock prices for stock time series data. A large number of works have been done based on ANN techniques for stock market prediction [12-20]. Many applications of hybrid systems in stock market time series data analysis have also been proposed in the literature [21-25].

In contrast to the work mentioned above, our approach in this paper is based on structural decomposition of a time series to study the behavior of the auto sector in India during 2010-2015. In addition, we have proposed five forecasting techniques, and one method to understand the contribution of the various constituents of a time series. We have computed the relative accuracies of each of the forecasting techniques, and also have critically analyzed under what situations a particular technique performs better than the other techniques. Our proposed framework of analysis can be used as a broad approach for forecasting the behavior of other stock market indices in India. Our results also elicit one point clearly – the auto sector index in India had a sharp change in trend during September – December 2015 that made forecasting task quite challenging if the forecast horizon is long.

## 7. CONCLUSION

In this work, we have analyzed the auto index time series in India during the period January 2010 to December 2015. We have used R programming language to structurally decompose the time series values into three components - trend, seasonal, and random. The decomposition results have provided us a deeper insight into the behavior of the auto index time series. Based on the results, we have been able to identify the months during which the seasonal component plays a major role. We have also been able to have an idea about the trend of the auto sector index. Using these decomposition results, we have proposed five approaches for forecasting the index values of the auto sector with a forecast horizon as large as 12 months. We have also introduced a technique to understand the structural analysis of the time series data using its trend and seasonal components. The forecast results clearly demonstrate the effectiveness and efficiency of our proposed forecasting techniques. Even in presence of a random component and a sharply changing trend values in time series values, our techniques have been able to achieve quite an acceptable level of forecasting accuracies.

The results obtained from the above analysis is extremely useful for portfolio construction. When we perform this analysis for other sectors as well, it will help portfolio managers and individual investors to identify which sector, and in turn which stock, to buy/sell in which period. It will also help in identifying which sector, and hence which stock, is dominated by the random component and thus is speculative in nature.